\documentclass[aps,twocolumn,prd,showpacs,amsmath,amssymb,nofootinbib]{revtex4-2}
\usepackage{dcolumn}
\usepackage{bm}
\usepackage{amsfonts}
\usepackage{amsthm}

\newcommand{\be}{\begin{equation}}
\newcommand{\ee}{\end{equation}} 
\newcommand{\eei}{\end{equation}\indent\indent}
\newcommand{\bc}{\begin{center}}
\newcommand{\ec}{\end{center}}
\newcommand{\ber}{\begin{eqnarray*}}
\newcommand{\ear}{\end{eqnarray*}}
\newcommand{\ba}{\begin{array}}
\newcommand{\ea}{\end{array}}

\newcommand{\bea}{\begin{eqnarray}}
\newcommand{\eea}{\end{eqnarray}}

\newcommand{\ei}{\end{itemize}}

\newcommand{\ud}{\dot{u}}

\newcommand{\bra}[1]{\left(#1\right)}

\newcommand{\A}{{\cal A}}

\def\case#1/#2{\textstyle\frac{#1}{#2} }

\def\Journal#1#2#3#4{{#4} {\it #1} {\bf #2}, #3 }

\def\ud{\textrm{d}}

\newcommand{\smfrac}[2]{{\textstyle{#1\over#2}}}

\newcommand{\w}[1]{\bm{#1}} 
\def\ud{\textrm{d}}
\begin{document}

\title{Comment on "Shear-free barotropic perfect fluids cannot rotate and expand simultaneously" by R.~Goswami and G.F.R.~Ellis}
\author{Norbert Van den Bergh}
 \email{norbert.vandenbergh@gmail.com}
  \affiliation{Department of Electronics and Information Systems, Faculty of Engineering and Architecture, Ghent University, Ghent, Belgium}
 \author{John Carminati}
 \email{jcarm1930@gmail.com}
 \affiliation{School of Information Technology, Deakin University, Australia}\date{\today}


\begin{abstract}
We point out an error in a recent paper by Goswami and Ellis. As a consequence the question of whether shear-free barotropic perfect fluids (with $p+\mu\neq 0$) can or cannot rotate and expand simultaneously, is still wide open.
 \end{abstract}
\pacs{}

\maketitle

\section{Introduction}

Recently there has been renewed interest in the so-called \emph{shear-free fluid conjecture}, which claims that
\begin{quote}
general relativistic, shear-free perfect fluids obeying a barotropic
equation of state such that $p + \mu \neq  0$, are either non-expanding ($\theta\neq 0$)
or non-rotating ($\omega\neq 0$).
\end{quote}
For an overview of the history of the problem we refer to \cite{VdB-Radu2016}, which contained as main results the following theorems: (1) the conjecture is correct for linear equations of state $p = a \mu + b$ ($a,b$ constants, thereby also incorporating the existence of a possible non-0 cosmological constant) and (2) if a shear-free perfect fluid with an general barotropic equation of state is non-rotating and non-expanding, then vorticity and acceleration are mutually orthogonal if and only if a Killing vector exists along the vorticity. Note that, while in the latter case
the equations describing the problem simplify dramatically, the accompanying loss of information turns this special case, as remarked already by
Collins \cite{CollinsKV}, into an exceptionally elusive one. While the search for a counter-example to the conjecture or the desire to construct a proof has lead to the investigation of a large number of --admittedly sometimes contrived\footnote{see for example the present authors' investigations of shear-free perfect fluids with a solenoidal electric or magnetic part of the Weyl curvature \cite{NorbertCarminatiKarimian2007} \cite{NorbertKarimianCarminatiHuf2012}}-- sub-cases (so far however all in agreement with the conjecture), it also has lead occasionally authors to let their cautiousness slip away, resulting in incorrect claims \footnote{see for example \cite{Carminati2015} where a bug in Maple's solve routine went unnoticed, or \cite{Sikhonde_Dunsby2017} where the abstract of the paper incorrectly claimed to provide a covariant proof for the case where the acceleration and vorticity vectors are orthogonal}. The recent publication \cite{Goswami_Ellis_grqc20021} by R.Goswami and G.F.R.~Ellis, claiming to have obtained a general proof of the conjecture, is a further example of the latter. In their paper the authors use the 1+1+2 formalism, which is a specification of the more well known local 1+3 threading of the space-time manifold, using the unit time-like unit vector $\w{e}_0= \w{u}$ along the fluid flow lines and the unit vector $\w{e}_3$ along the vorticity $\bm{\omega}$ as preferred vector fields. This formalism bears a strong similarity to the orthonormal tetrad formalism \cite{MacCallum1971} \cite{EllisMaartensmacCallum2012}, in so far that the dot and hat derivatives appearing in \cite{Goswami_Ellis_grqc20021} are identical (when acting on scalars $\phi$) to the directional derivatives $\w{e}_0, \w{e}_3$ ($\dot{\phi}=\w{e}_0(\phi), \hat{\phi}=\w{e}_3(\phi)$). 
On the other hand, the directional derivative operators $\w{e}_1,\w{e}_2$ can be re-defined by absorbing some of the "badly transforming spin coefficients" (under rotations in the (1,2) planes), so as to obtain the $\delta$ operators of the 1+1+2 formalism. For comparison of the equations appearing in both formalisms, we present in table \ref{table1} the correspondence between some of the relevant variables used in \cite{Goswami_Ellis_grqc20021} and their analogs in \cite{VdB-Radu2016}.
\begin{table*}[ht] \label{table1}
\caption{1+1+2 variables vs.~orthonormal tetrad variables, indices being ordered as (1,2,3,0)}
\centering
\begin{tabular}{p{0.35\linewidth}p{0.35\linewidth}}
\hline
\noalign{\vskip 2mm}
1+1+2 & orthonormal tetrad \\
\noalign{\vskip 2mm}
\hline
\noalign{\vskip 2mm}
$\Omega$ & $\omega_3$\\
$\A$ & $\dot{u}_3$\\
$\xi$ & $ n_{33}/2$\\
$\phi$ & $-2a_3$\\
$\alpha^a$ & $[\Omega_2,\, -\Omega_1, \, 0, \, 0]$\\
$\A^a$  & $[\dot{u}_1, \, \dot{u}_2, \, 0, \, 0]$  \\
${\it a}^a $ & $[a_1-n_{23}, \,a_2+n_{13}, \, 0, \, 0]$  \\
$\zeta_{ab}$ & $\left[
           \begin{array}{cccc}
             -n_{12} & (n_{11}-n_{22})/2 & 0 & 0 \\
              (n_{11}-n_{22})/2 & n_{12} & 0 & 0\\
             0 & 0 & -a_3 & 0\\
             0 & 0 & 0 & 0\\
           \end{array}
         \right]$\\
$N_{ab}$ & $\left[
           \begin{array}{cccc}
            1 & 0 & 0 & 0\\
            0 & 1 & 0 & 0\\
            0 & 0 & 0 & 0\\
             0 & 0 & 0 & 0\\
           \end{array}
         \right]$\\
         \noalign{\vskip 2mm}
         \hline
\end{tabular}
\end{table*}
We note the different uses of the symbols in the two formalisms: the vorticity scalar is represented by $\Omega$ in \cite{Goswami_Ellis_grqc20021} and by $\omega_3$ in \cite{VdB-Radu2016}, while the quantities $\Omega_1,\Omega_2,\Omega_3$ in the latter specify the rotation of the spatial triad with respect to a Fermi-propagated triad.\\
In Lemma 1 of \cite{Goswami_Ellis_grqc20021} it is proved that the projection of the time evolution of the unit vector along the vortex line onto the (1,2)-plane vanishes: $\alpha_a=0$. In the tetrad-approach this is reflected by the conditions $\Omega_1=\Omega_2=0$, which follow by applying the commutators $[\w{e}_3, \w{e}_1]$ and $[\w{e}_3, \w{e}_2]$ to $p$ and using the conservation laws. It is then customary to apply a rotation in the (1,2)-plane such that $\Omega_3+\omega_3=0$, ensuring that the spatial triad is Fermi-propagated along the flow. Note that this fixes the spatial triad only modulo \emph{basic} rotations, namely rotations for which the rotation angle $\psi$ satisfies $\dot{\psi}=0$: in \cite{VdB-Radu2016} such a rotation is used to fix the tetrad further such that $n_{11}-n_{22}=0$, therewith diagonalizing the matrix $\zeta_{ab}$. A further fixation making also $n_{12}=0$ is possible when a Killing vector exists along the vorticity.\\
A consequence of the vanishing of $\alpha_a$ is the following important set of "derivative equations", (87-92) of \cite{Goswami_Ellis_grqc20021}:
\bea
\dot{\Omega}&=&\left(c_s^2-\frac23\right)\Omega\theta\;,\label{Omdot}\\
\hat{\Omega}&=&(\A-\phi)\Omega\;,\label{Omhat}\\
\dot\xi&=&-\frac13\theta\xi\;,\label{xidot}\\
\dot{\phi}&=&\frac23\theta\left(\A-\frac12\phi\right)+2\xi\Omega\label{phidot},\\
\hat{\theta}&=&3\xi\Omega\;,\label{That}\\
\dot{\A}&=&-\A\theta\bra{\frac{1}{c_s^2}\frac{d^2p}{d\mu^2}(\mu+p)-c_s^2+\frac13}+3c_s^2\xi\Omega,
\label{Adot}
 \eea
where $c_s^2=\frac{\ud p}{\ud \mu}$. Note that these equations are obtained \emph{only} making use of the commutator relations, the conservation laws,
\bea
\dot{\mu}+(\mu+p) \theta=0,\label{mudot}\\
\w{e}_\alpha (p)+ (\mu+p) \dot{u}_\alpha =0\  (\alpha=1,2,3),\label{gradp}
\eea
\emph{and} the $(a=3)$ component of $C_1$, the constraint equation (30) of  \cite{Goswami_Ellis_grqc20021}:
it is easy to verify that they are identically satisfied under the commutation relations and conservation laws, with (\ref{xidot}) as single exception. The latter in addition requires\footnote{this will be relevant for the construction of the counter example below} as extra information the (03) Einstein field equation, or, equivalently, the 3d component of constraint equation $C_1$.

Using the above derivative equations the authors of \cite{Goswami_Ellis_grqc20021} show then in their Theorem 1 that for a rotating shear-free barotropic perfect fluid the flow lines and the vortex lines are 2-surface forming. This is correct and, in fact, one can use the commutator relation $[\w{e}_0,\, \w{e}_3]= \dot{u}_3 \w{e}_0-\smfrac{1}{3} \theta \w{e}_3$ to show that $[\lambda^3 \epsilon^{-1} \w{e}_0, \, \lambda^{-1} \w{e}_3]=0$, where $\lambda$ and $\epsilon$ are functions of the matter density defined by
$ \lambda = \exp \int \frac{\ud \mu}{3(p+\mu)}$, $\epsilon=\mu+p$: coordinates $t,z$ therefore exist such that $\w{e}_0=\epsilon \lambda^{-3} \frac{\partial}{\partial t}$ and $\w{e}_3=\lambda \frac{\partial}{\partial z}$. Adding coordinates $x,y$ such that
\be
\w{e}_{I} = \mathcal{A}_I \frac{\partial}{\partial t}+\mathcal{B}_I \frac{\partial}{\partial z}+\mathcal{C}_I \frac{\partial}{\partial x}+\mathcal{D}_I \frac{\partial}{\partial y},
\ee
$I=(1,2)$, these 2-surfaces are then just the $(x=constant, y= constant)$ surfaces. \\

The crucial step of \cite{Goswami_Ellis_grqc20021} comes in Part 1 of Theorem 2, where an argument is presented which claims to guarantee a factorisation (101) of the vorticity scalar,
\be
\Omega=\Omega_1\Omega_2\,\neq0,   \,\delta^a\Omega_1=0,\,\dot\Omega_2= \hat\Omega_2=0.
\label{Om12}
\ee
This is a remarkable claim, it being based on the use of the commutator relations, the conservation laws (\ref{mudot}, \ref{gradp}) and a single field equation only. It is even more remarkable, as this would solve the above mentioned "elusive case" with a single stroke of the pen: the existence of a non-constant function $\Omega_1=\Omega_1(t)$ \footnote{when a Killing vector along the vorticity exists, not only $\delta^a\Omega_1=0$ but also $\hat\Omega_1=0$} 
would immediately imply that $\w{u}$ is hypersurface-orthogonal, whereas for a constant $\Omega_1$, $\Omega$ would be \emph{basic} and hence $\theta$ would be 0 by (\ref{Omdot}). As we have been unable to convince the authors of their mistake, we proceed to present a counter-example.

\section{Counter-example}

Of course one cannot expect a counter-example to the conjecture itself, as then the whole issue would have been finished long ago. What we can do however, is construct an example illustrating the failure of the reasoning given in \cite{Goswami_Ellis_grqc20021}, by presenting an explicit space-time and a flow describing a shear-free barotropic perfect fluid for which all the equations (\ref{Omdot}-\ref{Adot}) are satisfied and where the fluid's energy-momentum tensor is also divergence-free, such that the conservation laws (\ref{mudot}, \ref{gradp}) hold. As the tricky point is to guarantee the validity of (\ref{xidot}), i.e.~of the $C_1$ constraint or the (03) field equation, we opt for an example in the "elusive situation" where the acceleration is orthogonal to the vorticity. In that case the hat-operator acting on all scalars is identically 0, implying $\A=\phi=\xi=0$ and we only have to take into account the dot-operators. As a toy-model we therefore take a space-time in which the orthonormal tetrad $\w{e}_0,\w{e}_1,\w{e}_2,\w{e}_3$ is defined by the dual basis,
\bea
\omega^0 &=& \smfrac{1}{2}\mu^{-1/2} (\ud t – y \ud x),\\
\omega^1 &=&  \mu^{-1/6} \ud x,\\
\omega^2 &=& \mu^{-1/6} \ud y,\\
\omega^3 &=& \mu^{-1/6} (\ud z – \ud x),
\eea
where $\mu=\mu(t,x,y)$ is arbitrary. Defining the velocity one-form to be $\omega^0$, it is an easy exercise to verify that the following "stiff fluid" energy momentum tensor,
\be
T_{ab}= \mu ( 2 u_{a}u_{b}+ g_{ab}),
\ee
is divergence-free, while the fluid's kinematical scalars are given by
\bea
\theta &=&- \mu^{-1/2} \mu_{,t},\\
\dot{\w{u}} &=& -\smfrac{1}{2}\mu^{-2/3}[(y\mu_{,t}+\mu_{,x}) (y \partial_t + \partial_x + \partial_z)+\mu_{,y}\partial_y],\\
\Omega &=& \smfrac{1}{2} \mu^{-1/6} 
\eea
and satisfy the full set (\ref{Omdot}-\ref{Adot}).
As $\Omega$ is clearly not factorizable for a general choice of the function $\mu$, this shows that at least \emph{some} extra argumentation will be needed --based on the full set of field equations and/or Bianchi equations-- in order to warrant the claim made in \cite{Goswami_Ellis_grqc20021}.
\newpage
{}
\vspace{0.2in}


\end{document}